
\input phyzzx
\FRONTPAGE
\line{\hfill BROWN-HET-927 and LPTENS-93-52}
\line{\hfill December 1993}
\vskip1.5truein
\titlestyle{{LOOP SPACE HAMILTONIANS AND FIELD THEORY OF NON-CRITICAL STRINGS}}
\foot{Work supported in
part by the Department of Energy under
contract DE-FG02-91ER40688-Task A}
\author{Antal JEVICKI}
\centerline{{\it  Laboratoire de Physique Th\'eorique}}
\centerline{{ Ecole
Normale Sup\'erieure, 24 rue Lhomond, 75230 PARIS Cedex 05, FRANCE.}}
\centerline{{\it and Department of Physics}}
\centerline{{\it Brown University, Providence, RI 02912, USA}}

\vskip .15in
\centerline{{\it and}}
\vskip .15in
\author{Joao RODRIGUES}
\centerline{{\it Physics Department}}
\centerline{{\it University of Witwatersrand, Johannesburg, South Africa}}
\bigskip
\abstract
We consider the loop space representation of multi-matrix models. Explaining
the origin of
a time variable through stochastic quantization we make contact with
recent proposals of Ishibashi and Kawai. We demonstrate how collective
field theory with its loop space interactions generates a field theory of
non-critical strings.
\endpage

\centerline{{\bf 1.  Introduction}}

Recent studies of matrix models have given beautiful solutions of non-critical
string theory
in low dimensions $9) \leq d \leq 1)$ [1]. A significant new insight is
contained in the appearance
of $W_k$ type symmetries which are most clearly exhibited in the
Schwinger-Dyson approach [2]. At $d=1$ one has a Hamiltonian problem with
physical
time representing $c=1$ matter and a field theoretical formulation with a
$W_{\infty}$
spectrum-generating algebra  provided by collective field theory [3].

Field-theoretic formulations are also
 of interest for $d <1$
[4,5]. Recently a Hamiltonian method was also suggested for $d <1$ [6] based on
a
time variable inferred from studies of $2d$ continuum gravity [7]. For $d=0$
one recognizes
that the Hamiltonian
is identical in structure to the one matrix collective field hamiltonian, It
contains simple
processes of joining and splitting of strings, given by a cubic interaction.

In what follows we will consider the matrix model approach further,
concentrating on chain integrals.  We will
present an interpretation of the time variable as originating from stochastic
quantization [8,9].
Here the time is introduced to give the evolution of probability density and
the correlation functions
are then obtained in the limit $t \rightarrow \infty$. One is then lead to
Fokker Planck Hamiltonians
whose specific form is $$ H = - ( {\partial \over \partial X} - {\partial S
\over \partial X} ) {\partial
\over \partial X}\eqno\eq $$.

Representing these Hamiltonians in loop space through standard collective field
formalism [10]
leads us to field theories of non-critical strings.
The main feature of such loop space Hamiltonians is a cubic interaction and
a consistency condition coming from
a requirement of hermiticity. Tha latter leads to a system of
equations obeyed by the partition function. These are known in matrix models as
$\tau$ function
conditions.

In general, already for two matrices the loop space becomes very large.
For specific critical potentials however we can consistently truncate loop
space to a finite number of
string fields. The consistency of this trunctaion is assured by closure of the
corresponding algebra.

The content of the paper is as follows. In Section 2 we summarize the basics of
the Schwinger-Dyson equations
and collective field representation.In Section 3 we describe (through
stochastic
quantization) the origin of time variables in the problem of $d <1$ matrix
models, and
introduce the corresponding Hamiltonians. Their loop space representation is
then discussed.
It is shown in Section 4 how a hermiticity requirement implies the constraint
 equations on the partition function.
In Section 5 we discuss in detail the $1$-matrix example and its simple
extensions. Particular
attention is given to taking the continuum limit and establishing contact with
[6]. In Section 6
we consider the problem of loop space Hamiltonians in the two-matrix case. It
is shown
 how a simple truncation leads to field
theories with a finite number of fields.

\vskip .10in

\centerline{{\bf 2.  Schwinger-Dyson Equations and Collective Field
Integral}}

\vskip .10in

Consider a partition function given by the integral
$$Z = \int dM_1 \cdots dM_k e^{-S}\eqno\eq$$
with hermitian $N\times N$ matrices $M_i$, $i=1, \cdots, k$ and an
action
$$S = -c \sum_i \, Tr (M_i M_{i+1} ) + \sum_i \, V_i (M_i )$$
The correlation functions in general obey the Schwinger-Dyson
equations which follow from
$$\int [dM] {d\over dM} \left( e^{-S(M)} F(M)\right) = 0\eqno\eq$$

The $U(N)$ invariant observables are given by traces ($\phi_n = \Tr
(M^n)$) for a single matrix and
$$\phi_C = \Tr (M_1^{n_{1}} M_2^{n_{2}} \cdots )\eqno\eq$$ in general.

The loop space index $C$ in general denotes the sequence of matrices
in trace $C = \{ n_1 , n_2 , \cdots  \}.$ In Yang-Mills
theories they are Wilson loops $W(C) = Tr \, e^{\int A}$

Here too we can use other parametrizations of invariant variables, for instance
$\Phi (l)
=Tr (e^{-Ml})$, and generalizations thereof.

The invariant S-D equations are obtained by contracting $U(N)$
indices and are
$$\langle - {\partial S\over\partial M_{\alpha} F^{\alpha}}  +
{\partial F^{\alpha}\over \partial M_{\alpha}} \rangle =
0\eqno\eq$$

Collective field theory represents a reformulation of the theory (in
this case the matrix integral) in terms of invariant loop variables
$\phi_C$.  The partition function is written as an integral
$$Z = \int \left[ d\phi_c\right] J (\phi ) e^{-S}\eqno\eq$$
Here the  new element is a nontrivial Jacobian coming from a
change of variables $M_i = t_{\alpha} M^{\alpha} (i) \rightarrow
\phi_c$.  The Jacobian was discussed in general in [10],
it is specified by a set of differential equations
$$\left[\Omega (c,c^{\prime} ) {d\over d\phi_{c'}} + \omega (\phi_c) +
{d\Omega (c,c^{\prime})\over d\phi_{c'}} \right] J = 0\eqno\eq$$
Here the repeated indices $(C')$ are summed over.
The main ingredients in these are the quantities $\Omega$ and $\omega$.
They are linear and quadratic (in the field $\phi$) respectively.
One has
$$\Omega (c,c^{\prime}) \equiv {\partial\phi_c\over\partial
M_{\hat{\alpha}}} \, {\partial \phi_{c'}\over \partial
M_{\hat{\alpha}}} \eqno\eq$$
where $\hat{\alpha} = (\alpha , i)$ denotes a sum over both $U(N)$ and
other matrix indices.  Using an identity
$$\sum_{\alpha} Tr (A t_{\alpha} ) Tr (Bt_{\alpha}) = Tr (AB)\eqno\eq$$
for a complete set of $U(N)$ generators one finds that
$$\Omega (c,c^{\prime} ) = \sum \phi_{c+c'}\eqno\eq$$
representing a joining of loops $c$ and $c^{\prime}$. The sum is over
all possible distinct joinings.  Similarly
$$\omega(c) \equiv - {\partial^2 \phi_c\over \partial M_{\hat{\alpha}}
\partial M_{\hat{\alpha}}} = \Sigma \phi_{c'} \phi_{c''}\eqno\eq$$
represents the process of splitting (of a loop $C$ into $C^{\prime},
C^{\prime\prime})$. Again, in general there can be several different
ways to split a loop. With the Jacobian specified by the loop joining
and splitting processes (eqs 9-10).
we can show that the collective field theory representation
generates the correct matrix model Schwinger-Dyson equations.
 Consider the following
total derivative in the loop space integral representation:
$$\int \left[ d\phi \right] {d\over d\phi_{C'}} \left( \Omega
(c,c^{\prime} ) J \, e^{-s} \prod_i f_i \right) = 0 \eqno\eq$$
Using the identity for the Jacobian this is shown to give
$$\langle \left[ \Omega (c,c^{\prime} ) {d\over d\phi_{c'}} -
\omega (c) - \Omega (c, S)\right] \prod_i f_i \rangle =
0\eqno\eq$$
with
$$\Omega(c^{\prime},S) = {\partial\phi_c\over
\partial M_{\hat{\alpha}}} \, {\partial S\over\partial
M_{\hat{\alpha}}}\eqno\eq$$
we recognize in the above the S-D equation for $F_{\alpha} =
{\partial\phi_c\over\partial M_{\alpha}} \, \prod f_i$

Clearly the collective Jacobian $J$ is determined in such a way
that the correct S-D equations are generated.
We comment that the above demonstration extends
to higher genuses the arguments  given in [10]
where the planar limit was considered.

To illustrate the above with the single matrix example one has
$$\eqalign{ \phi_n & = \Tr (M^n )\cr
 \Omega (n,n^{\prime})& = nn^{\prime} \phi_{n+n'-2}\cr
 \omega (n) & = -n \sum_{n'=0}^{n-2} \, \phi_{n'} \phi_{n-n'-2}
}\eqno\eq$$
with the resulting S-D equation
$$\langle \left[ \sum_{m=n}^{\infty} \phi_{n+m} m {d\over d\phi_m}
+ \sum_{m=0}^n \phi_m \phi_{n-m} + \phi_{n+m} m j_m \right]
F\rangle = 0\eqno\eq$$
for the general action $S= j_n \phi_n$.

\medskip
\centerline{{\bf 3.  Hamiltonians}}
\smallskip
The
relevance of Hamiltonians  in evaluating euclidean
correlation functions comes in general through the stochastic quantization.
For a typical field theory $\varphi (x)$ the time dependent
Langevin equation is given by
$${\partial\over\partial t} \varphi (n,t) = - {\partial S\over\partial\varphi}
+ \eta\eqno\eq$$
where $\eta$ is the random variable.  The correlation functions are then
obtained in the limit $t \rightarrow \infty$
$$\langle F(\varphi )\rangle = \lim_{ t\rightarrow\infty}
\int \left[ d\varphi (x) \right] F(\varphi ) P_t\eqno\eq$$
where the time evolution (of the probability distribution) is given
by
 $${\partial\over \partial t} P_t = - H_{FP} P_t\eqno\eq$$
with the Fokker-Planck Hamiltonian
$$H_{FP} = -{1\over 2} \int \left( {d\over d\varphi (x)} - {\delta S\over
\delta \varphi (x)} \right) {d\over
d\varphi(x)}\eqno\eq$$
This Hamiltonian has the obvious property that it is made Hermitian by a
similarity
transformation
$$e^{-1/2S} H_{FP} \, e^{1/2S} = - {1\over 2} \int \left( {d\over
d\varphi} - {1\over 2} S^{(1)}\right) \left( {d\over d\varphi} +
{1\over 2} S^{(1)}\right)\eqno\eq$$

This hermitian form guarantees that the Hamiltonian has a unique ground
state given by the wavefunction
$$\psi_0 = e^{-1/2 S}\eqno\eq$$
The usefulness of stochastic quantization for loop equations
in Yang-Mills theoryhas been pointed out by Marchesini [8]
(it was considered in detail in the one
matrix case in [9] with the purpose of numerical solutions of
S-D equations).
For the specific case of matrix models the formulation of the
Hamiltonian goes as follows.  One has
$$H_{FP} = -\left( {\partial\over\partial M_{\hat{\alpha}}} - {\partial
S\over \partial M_{\hat{\alpha}}} \right) {\partial\over \partial
M_{\hat{\alpha}}}\eqno\eq$$
as a nonhermitian matrix model hamiltonian.  Here $\hat{\alpha} =
(\alpha , i)$ represents a summation over $U(N)$ and other matrix
indices.  The change to loop space variables $\phi_c$ is then done,
in a  standard collective field theory fashion. The corresponding loop space
Hamiltonian
reads
$$H = - \left( \Omega (c, c^{\prime} ) {d\over d\phi_{c'}} - \omega
(c) - \Omega (c,S)\right) {d\over d\phi_c} \eqno\eq$$
with a summation over the repeated loop indices $c,c^{\prime}$ and the
loop joining and splitting quantities $\Omega$ and $\omega$ were given
earlier.  The additional term  $\Omega (c,S) \equiv {\partial\phi_c\over
\partial M} {\partial S\over\partial M}$ introduces the specific
matrix model action whose correlation functions are under
consideration.  The above Hamiltonian is cubic in the field $\Phi_c$
and its conjugate $\Pi_c = {d\over d\Phi_c}$ with the standard
canonical commutation relations
$$\left[ \phi_c, \Pi_{c'} \right] = - \delta_{c,c^{\prime}}\eqno\eq$$
It is useful to write the Hamiltonian in the notation
$$H = - \sum_c \hat{O}_c {d\over d\phi_c}\eqno\eq$$
with the differential operator
$$\hat{O}_c = \sum \Omega (c, c^{\prime}) \Pi_{c'} - \tilde{\omega}
(c)\eqno\eq$$
where in the $\tilde{\omega}$ we have added the term coming from the
action.

In general
the closure of the algebra generated by the operators $\hat O_c$ will be of
central relevance. It represents a central consistency condition in the
collective field formalism
\medskip
\centerline{{\bf 4  Hermiticity Condition and the partition-function}}
\smallskip
The loop space hamiltonian given above is seemingly nonhermitian, this
feature coming from the nonhermiticity of the standard collective
hamiltonian plus  the additional nonhermitian term introduced by the
action $S$.  It is a consistency requirement for the present
Hamiltonian approach that there should be a similarity transformation making
the
hamiltonian hermitian and implying the existence of unique ground
state.  This (unique) ground state wavefunction will be argued to
correspond to the $\tau$-function of matrix models.  The Virasoro and
the higher $W$ constraints then follow from the Hermiticity
requirement.

The manifestly hermitian Hamiltonian comes from the transformation [10]
$$\hat{H} = J^{-1/2} H J^{1/2}\eqno\eq$$
with the Jacobian obeying the equation
$$O_c^\dagger J \equiv \left( -{d\over d\phi_{c'}} \Omega (c,c^{\prime} ) -
\tilde{\omega} (c) \right) J (\phi ) = 0\eqno\eq$$
Using this equation one has a manifestly hermitian form
$$\hat{H} = \sum_{c,c'} \tilde{O}_c \Omega^{-1} (c,c^{\prime} )
\tilde O^\dagger_{c'}\eqno\eq$$
with
$$\eqalign{\tilde{O}_c & = \Omega (c, c^{\prime} ) {d\over d\phi_{c'}}
- {1\over 2} \tilde{\omega} (c)\cr
\tilde{O}_c^\dagger & = - {d\over d\phi_{c'}} \Omega (c,c^{\prime}) -
{1\over
2} \tilde{\omega} (c) }\eqno\eq$$
It is these operators that in the continuum limit of a particular
model give the conditions obeyed by the $\tau$-function.
This is seen as follows: Recall that
the partition function is given by
$$Z = \int \left[ d\phi\right] J (\phi) e^{-j_{c} \phi_{c} -
S}\eqno\eq$$
where we have in the action general couplings $j_c\phi_c$.
Since any terms in $S$ only redefines the couplings one essentially
has
$$Z (j) = \tilde{J} (j) = \int [d\phi ] J (\phi ) e^{-j \cdot
\phi}\eqno\eq$$
which says that the partition function is Laplace transform of the
Jacobian.  From the Hermiticity requirement it then follows that the square
root of the Jacobian obeys the constraint equations:
$$\tilde{O}_c^\dagger J^{1/2} = 0\eqno\eq$$
Denoting $J^{1/2} = \tau (j)$ we then have
$$\{ \sum {d \over d \Phi_{C'}} \Omega(C,C') -1/2 \tilde{\omega} (C) \} \tau =
0\eqno\eq$$
This is the statement that the $\tau$-functions (or the square root of
the partition functions) obeys a set of differential equations given by
the operators $\tilde{O}_c$.  As a concluding feature of this general
discussion let us comment further on the algebra of operators
$\tilde{O}^{\dagger}_c$.  It is a consistency condition for the
integrability of the above equations that the operators $O_c$ should
close in the sense that
$$\left[ \tilde{O}_c^\dagger , \tilde{O}_{c'}^\dagger \right] =
f_{ cc^{\prime}
c^{\prime\prime}} (\phi ) \tilde{O}_{c''}^\dagger \eqno\eq$$
Here in general one allows for field dependent structure constants.
Also in general the operators $O_c$ given by the collective
construction when commuted in principle can lead to further
generators
and in the above we mean the complete set. One can in turn
define generalized collective Hamiltonians based on closed algebras. We shall
 give simple examples of the latter at the end of next section, where
we  present some simple
extensions to the supersymetric case and the case of $SL(2,R)$
Kac Moody algebra.
\medskip
\centerline{{\bf 5.  Simple Examples}}
\smallskip
The one matrix collective field Hamiltonian has been extensively
discussed in the literature.  We begin our discussion with this
example, however, in order to demonstrate in detail the steps involved in
taking its
continum limit to $c=0$ theory. Previously this Hamiltonian was used for
studying
the $c=1$ theory in which case a simple $M^2$ potential is used.
The Hamiltonian appropriate for the $c=0$ integral with the action
$$S = Tr \left( {1\over 2} \mu M^2 - {1\over 3} M^3 + \cdots \right)\eqno\eq$$
reads
$$H = - Tr \left( {\partial\over\partial M} - {\partial S\over
 \partial M}\right) \, {\partial\over\partial M}\eqno\eq$$

Changing to the (Loop Space) Collective Field representation with
$$\phi_n = Tr \left( M^n \right) \, \, , \, \, \Pi_n = {\partial
\over \partial \phi_n} \,\,; \,\, n \geq 0\eqno\eq$$
The Hamiltonian becomes
$$H_x - \sum_n \left( \sum_{m=0}^{\infty} \phi_{n+m-2} m {\partial
\over\partial \phi_m} + \sum_{r=0}^{n-2} \phi_r \phi_{n-r} -
 \Omega \left( S, \phi_n \right) \right) n {\partial\over \partial
\phi_n}\eqno\eq$$
with
$$\Omega (S,\phi_n ) = n \left( \mu \phi_{n+1} - \phi_{n+2} + \cdots
\right)\eqno\eq$$
The source ($S$ independent) term is a cubic interaction.  It is
written as
$$H_3 = - \sum_n \, O_n \, n\Pi_n \eqno\eq$$
with
$$O_{n+2} = \sum_{n=0}^{\infty} \, \phi_{n+m} m \Pi_m + \sum_r \phi_r
\phi_{m-r}
\eqno\eq$$
and we have a Virasoro algebra $O_{n+2} = L_n$:
$$\left[ L_n , L_m \right] = \left( n-m\right) L_{n+m}\eqno\eq$$
The Hamiltonian couples the Virasoro generators to the
 conjugate field $\Pi_n$.  For evaluating the
continuum limit it is useful to switch to $z$-representation with
$$\phi (z) = Tr {1\over z-M} = \int_0^{\infty} dLe^{-Lz} \phi_L =
\int_0^{\infty} dLe^{-Lz} Tr (e^{LM})\eqno\eq$$
so that
$$\eqalign{\phi(z)& = \sum_{n\geq 0} \, z^{-n-1} \phi_n\cr
\partial \Pi (z) & =\sum_{n\geq 0} \, z^{n-1} n \Pi_n }\eqno\eq$$
with
$$O(z) =\sum_n \, z^{-n-2} O_n\eqno\eq$$
gives
$$O(z) = \int dz^{\prime} {\phi(z') - \phi (z)\over z - z'}
 \, \partial_{z'} \Pi + \phi^2 (z) \equiv (\phi \partial_{z}\Pi )(z) + \phi^2
(z)\eqno\eq$$
The bracket notation is to state that there are only $z^{-n}$
 components.  The Hamiltonian is now
$$H = - \int dz \left[ \phi (z) \partial_z \Pi (z) + \phi^2
(z) + \left( z^2 - \mu z\right) \phi + \left( \mu - z
\right) + c_0 \right] \, \partial_z \Pi (z) .\eqno\eq$$
The scaling limit is defined with
$$\eqalign{ \mu & = \mu_c + a^2 \Lambda\cr
z & = z_a + a \zeta } \eqno\eq$$
and a shift
$$\phi (z) = {1\over 2} (z\mu - z^2 ) + a^{3/2} \Phi
(\zeta)\eqno\eq$$
After the shift the linear $\phi$-term in the operator $O(z)$ gets
canceled and the complete potential term reads
$$ a^3 \left( \Phi^2 (\zeta ) - {1\over a^3} \left[
{1\over 4} \left( \mu z - z^2 \right)^2 + (z-\mu ) +
c_0 \right] \right) = a^3 \left( \Phi^2 (\zeta ) - \Phi_0^2 (z)
\right)\eqno\eq$$
where we have recognized the planar value of $\Phi^2 $.
 For the conjugate field
we have
$$\Pi (z) = a^{-5/2} \Pi (\zeta)\eqno\eq$$
To summarize in terms of the basic length $a$,
 the dimensions are $[\zeta ] = a^{-1}$, $[\Phi (\zeta) ]
 = a^{3/2}$ and $[\Pi (\zeta) ] = a^{-5/2}$.  If one transforms

$$\Phi (\zeta ) = \int_0^{\infty} dl \,\,e^{-l\zeta} \Phi
(l)\eqno\eq$$
to the loop space $\Phi (l) , \Pi (l)$ fields,
 the dimensions are:
$$[l] = a \,\, [\Phi (l) ] = a^{-5/2} \,\, [\Pi (l) ]
 = a^{3/2}\eqno\eq$$
Let us now turn to the kinetic term. First we exhibit $N$ by scaling
 $z\rightarrow \sqrt{N} z, \,\, \phi \rightarrow \sqrt{N}$ .
This leads to a factor $N^{-2}$ in front of the kinetic term $\int \phi
\Pi_z^2$.
Inserting the continuum fields we find that the shift does not contribute and
$$H = - a^{1/2} \int d\zeta \left[ {1\over N^2 a^5} \, \Phi
 \partial_{\zeta} \Pi + \Phi (\zeta )^2 - \Phi_0^2 \right]
 \partial \Pi (\zeta )\eqno\eq$$
In the first term we have the correct scaling dimension for
the coupling constant $[g] = a^{-5}$.  We also find an
overall $a^{1/2}$ which is removed by a redefinition of the
(stochastic) time.  Consequently the time has dimension
$a^{-1/2}$.  The final continuum Hamiltonian written for
the loop fields reads
$$ \eqalign{H &= - \int_0^{\infty} dl_1 \int_0^{\infty} dl_2 \left\{ g
\Phi \left( l_1 + l_2 \right) l_1 \Pi (l_1 ) l_2 \Pi (l_2 ) +
 \left( l_1 +l_2 \right) \phi (l_1) \Phi (l_2) \Pi (l_1 + l_2 )
 \right\} \cr &- \int_0^{\infty} dl\,\, \rho_0 (l) \Pi (l)} \eqno\eq$$
This indeed is identical to the form guessed in [6]. As we have demonstrated,
 the terms with the correct scaling survive in the continuum.
It is a fact that the cubic interaction
remained unchanged. This will be a general feature in all theories.

Let us now turn to some simple generalizations.
Since in the above we had a Virasoro generator $O(z) = L(z) =\phi
\partial \pi + \phi^2$ coupled to the conjugate $\pi(z)$ with
consistency assured by closure of the Virasoro algebra
we can contemplate simple extensions by using other algebras.

(i)  Supersymmetric $N-S \, (R)$ algebra.  Consider
$$\eqalign{L_n & = {1\over 2} \Sigma^{\prime} \alpha_{n+m}
 \alpha_{-m} + {1\over 2} \sum_r \left( r + {1\over 2}
 n \right) b_{n+m} b_{-n}\cr
G_r & = \sum_r \alpha_{-n} b_{r+n} }\eqno\eq$$
denote the conjugate pairs as
$$\eqalign{ \alpha_n = \phi_n , \alpha_{-n} = n \Pi_n &\,\,n>0\cr
b_r, b_{-r} = P_r &r>0 }\eqno\eq$$
then a natural form is
$$H = - \left ( \sum_{n>0} L_{n-2} m \Pi_n - {1\over 2}
\sum_r \, G_r \, rP_r \right)\eqno\eq$$

(ii)  $SL(2,R)$ Kac-Moody algebra.  One has the representation
 for the currents:
$$\eqalign{ J^+ (z) & = - \gamma^2\beta + i \alpha_+ \gamma
 \varphi + \kappa \partial\gamma\cr
J^0 (z) & = \beta \gamma - {i\over 2} \alpha_+ \partial
\varphi\cr
J (z)& = \beta }\eqno\eq$$
Denoting
$$\eqalign{ \partial\varphi (z) &+ \phi (z) + \partial_z \Pi (z)\cr
\beta (z) & = \sum_n \, z^{-n-1} \, \beta_n \,\, ; \,\, \beta_n =
{\partial\over\partial \gamma_{-n}} }\eqno\eq$$
 a collective type Hamiltonian reads:
$$H = \int dz \left\{ J^+ (z) \beta (z) - J^0 (z) \partial_z \Pi (z)
\right\}\eqno\eq$$

{\bf 6.  Two-matrix Hamiltonian}

It is known [11] that the whole sequence of non-critical string models with
increasing dimensions
$d_k = 1- {6 \over (k+1)(k+2)} , k = 1 \cdots n $ ,can be generated from
critical points of
the two matrix model. Denoting the two matrices as $M_1=X$ and $M_2= Y$, one
has the action.

$$
S = Tr (-c XY + v_1(X) + v_2(Y))
\eqno\eq$$

The first non-trivial $d=1/2$ non critical string theory corresponds to a cubic
potential:

$$
v(X) = X^2 /2 - \lambda / 3 X^3
\eqno\eq$$

while the higher models are generated by potentials of higher power in $X$ and
$Y$. We shall
concentrate our discussion on the first non trivial cubic case (i.e. $d=1/2$).
Generalisation to
arbitrary $d_k \leq 1$ or $k=3,4 \cdots$ will be straightforward.

The Hamiltonian operator for two matrices reads $$H= -Tr  (({\partial \over
\partial X}
-{\partial S\over\partial X}){\partial\over\partial X} +
({\partial\over\partial Y}
-{\partial S\over\partial Y}){\partial\over\partial Y}) \eqno\eq$$

As we have explained in the general discussion the information on the specific
potential comes in
through the source term $$ \hat{S} = -Tr (({\partial S\over\partial X}{
\partial\over\partial X}
+{\partial S\over\partial Y}{ \partial\over\partial Y}) \eqno\eq$$
which is
added to the two-matrix Laplacian operator ${\partial ^2\over\partial X^2}
+{\partial ^2
\over\partial X^2}$. In the transition to the collective field theory it is the
Laplacian which
produces the cubic loop space interaction while $\hat{S}$ in general
only adds a loop space source term.
The loop space variables are now given by the traces $$ \Phi_C = Tr (X^{n_1}
Y^{n_2}X^{n'_1}
Y^{n'_2} \cdots ) \eqno\eq$$ with  $C = \{ n_1 , n_2 , n'_1 , n'_2 \cdots \}$
denoting the set of
looops around two fixed points. This represents an extremely large space.
However it is known
from study of Schwinger-Dyson equations of the two matrix problem [12] that for
specific (polynomial) potentials the space reduces and the S-D equations  close
within a subset of
loops. For the first non-trivial case of a cubic potential we have two sets of
loops
$$ \Phi^0 _n = Tr X^n , n=0,1 ...\eqno\eq$$
$$ \Phi^1 _n = Tr X^n Y, n=0,1 ...\eqno\eq$$

We shall now show that on this subset one can define a consistent Hamiltonian.
In general as the
order of the potential $v_2(Y)$  increases to $ Y^{k+1}$
one will define a reduced theory with
$k$ independent fields.

We now evaluate the joining and splitting interactions specified by $ \Omega
(C,C')$
and $\omega (C)$. First:

$$ \Omega ( \Phi^0_n , \Phi^0_m ) = Tr ({\partial \Phi^0_n\over\partial X}
{\partial \Phi ^0 _m\over\partial X}) = nm \Phi^0_{n+m} \eqno\eq$$
$$  \Omega ( \Phi^0_n , \Phi^1_m ) = Tr ({\partial \Phi^0_n\over \partial X}
{\partial \Phi^1_m\over \partial X}) = nm \Phi^1_{n+m} \eqno\eq$$

Here loops of type $0$ join amongst themselves while a loop of type $1$
joins with a loop of type $0$ into a loop of type $1$.
Similarly one can evaluate
$ \Omega ( \Phi^1_n , \Phi^1_m)$.
Here one encounters new loops of the type $Y^2$. We will be truncating the
system and will specify
$\Omega ( \Phi^1_n , \Phi^1_m )$ in such a way as to achieve consistency (in
the reduced S-D
equations this quantity does not participate).

For the splitting interaction one simply has
$$\omega (\Phi^0_n) = n \sum_{r=0}^{n-2} \Phi^0_n \Phi ^0_{n-r-2} \eqno\eq$$
$$\omega (\Phi^1_n) = n \sum_{r=0}^{n-2} \Phi^0_n \Phi ^1_{n-r-2} \eqno\eq$$
 Here loops of type $0$ split as in the one-matrix case
while a loop of type $1$ splits into $0$ and $1$.
Next we evaluate the term induced by the potential

$$ \hat{S} = -\Omega(S, \Phi^0_n ){\partial\over \partial \Phi^0_n}
 -\Omega(S, \Phi^1_n ){\partial\over \partial \Phi^1_n}\eqno\eq$$

with

$$ -\Omega(S, \Phi^0_n ) = n(c \Phi^1_{n-1} - \Phi^0_n + \lambda \Phi^0_{n+1})
\eqno\eq$$
$$ -\Omega(S, \Phi^1_n ) = n(c/ \lambda (\Phi^1_{n-1} - c\Phi^0_n) - \Phi^1_n +
\lambda \Phi^1_{n+1}) \eqno\eq$$

In this last expression the first term comes from
eliminating a higher loop variable $\Phi^2_m$. The substitution is justified by
the
S-D equation :

$${\partial S\over \partial Y} {\partial \Phi^1_m\over \partial Y}
 -{\partial \over \partial Y} {\partial \Phi^1_m\over \partial Y} = 0\eqno\eq
$$

which gives

$$ \Phi^2_m = 1/ \lambda (\Phi^1_{m-1} - c\Phi^0_m) \eqno\eq $$

We now have all the ingredients to define the reduced Hamiltonian. The fields
are now coupled to
the two operators :

$$O^0_n = \Phi^0_{n+m-2} m {d\over d \Phi^0_m} + \Phi^0_m \Phi^0_{n+m-2}
+\Phi^1_{n+m-2} m {d\over d \Phi^1_m} + \Omega(S, \Phi^0_n ) \eqno\eq $$
$$O^1_n = \Phi^1_{n+m-2} m {d\over d \Phi^1_m} + \Phi^1_m \Phi^0_{n+m-2}
+\Phi^1_{n+m-2} m {d\over d \Phi^0_m} + \Omega(S, \Phi^1_n ) \eqno\eq $$

As predicted by the general theory one can see that:

$$<O^{(0),+}_n \Pi_{i} \Phi^0_{n_i})> = 0 \eqno\eq$$
$$<O^{(1),+}_n \Pi_{i} \Phi^0_{n_i})> = 0 \eqno\eq$$

give the correct S-D equations. Indeed these equations read :

$$< (c\Phi^1_{n-1} - \Phi^0_n + \lambda \Phi^0_{n+1} + \sum \Phi^0_r
\Phi^0_{n+r-2}) \Pi_i
\Phi^0_{n_i}> + \sum_{i} n_i < \Phi^0_{n+n_i -2}\Pi_{j \neq i} \Phi^0_{n_j} >
=0 \eqno\eq$$
$$ \eqalign{ < (c / \lambda (\Phi^1_{n-1} - c\Phi^0_n) - \Phi^1_n &+
\lambda \Phi^1_{n+1} + \sum \Phi^1_r \Phi^0_{n+r-2} \Pi_i \Phi^0_{n_i} > \cr &+
\sum_{i} n_i < \Phi^1_{n+n_i -2}\Pi_{j \neq i} \Phi^0_{n_j} >  =0 }\eqno\eq$$

which can be compared with [12].

Continuing with the Hamiltonian, it is given as :

$$H = \int dz (O^0 (z) \Pi_0 (z) + O^1 (z) \Pi_1 (z) \eqno\eq $$

where we have the fields

$$ \Phi^i (z) = \sum_{n >0} z^{-n-1} \Phi^i_n , i=0,1 \eqno\eq $$
$$ \Pi^i (z) = \sum_{n >0} z^{n-1} n {\partial\over\partial \Phi^i_n} , i=0,1
\eqno\eq $$

Explicitely:

$$  \eqalign{ H &= - \int  {\bf [} \Phi^0 (z) \Pi^0 (z) + (\Phi^0 (z))^2 +
\Phi^1 (z) \Pi^1 (z)
 {\bf ]}\Pi^0 (z) \cr
&- \int {\bf [}\Phi^0 (z) \Pi^1 (z) + \Phi^1 (z) \Pi^0 (z) +\Phi^1 (z) \Phi^0
(z){\bf ]} \Pi^1 (z)
+ \hat{S}(z) } \eqno\eq$$

Let us now elaborate on the consistency of the truncation followed to reach
this
Hamiltonian. As we have explained, a consistent collective field theory
requires a
closure of the operator algebra $ \{ O_i \}$. Here we can easily see that this
holds.
Shifting $\Pi_1 \rightarrow \Pi_1 + \Phi_1$ and defining conformal fields :
$$ \partial \phi_0 = \sqrt{2} \Phi_0 + 1/\sqrt{2} \Pi_0 \eqno\eq$$
$$ \partial \phi_1 = \sqrt{2} \Phi_1 + 1/\sqrt{2} \Pi_1 \eqno\eq$$

we have
$$O_0 (z) = 1/2 ((\partial \phi_0)^2 + (\partial \phi_1)^2)_{-} \eqno\eq$$
$$O_1 (z) = (\partial \phi_0 \partial \phi_1) \eqno\eq$$

These operators close the algebra
$$ [O^0_n, O^0_m ] = (n-m) O^0_{n+m} \eqno\eq $$
$$ [O^0_n, O^1_m ] = (n-m) O^1_{n+m} \eqno\eq $$
$$ [O^1_n, O^1_m ] = (n-m) O^0_{n+m} \eqno\eq $$
Clearly this can be decoupled by setting $ O^{\pm} \equiv O^0 \pm O^1$ into
two decoupled Virasoro algebras.

Introducing the linear combinations

$$\Phi_{\pm} = \Phi^0 \pm \Phi^1 \eqno\eq$$
$$\Pi_{\pm} =1/2  \Pi^0 \pm \Pi^1 \eqno\eq$$

the Hamiltonian is written in a more symmetric form as:

$$ \eqalign{ H= &- \int [ 2 \Phi_+ \Pi_+ + 1/2 \Phi_+ \Phi_+ + 1/2 \Phi_+
\Phi_- ] \Pi_+
\cr &+ \int
[2 \Phi_- \Pi_- + 1/2 \Phi_- \Phi_- + 1/2 \Phi_+ \Phi_- ] \Pi_- + \hat{S}(z) \}
}\eqno\eq$$

In the continuum limit one expects that the non-scaling term disappears
The cubic interaction terms certainly survive; the fields now have the
dimension
$[\Phi(\zeta)] = a^{-4/3} , [ \Pi ] = a^{10/3} $.
As always the coupling constant $g$ can be introduced into the Hamiltonian
 by a constant rescaling
 of the fields.

 If one expands, taking the following:

$$\Phi_+ (z) = \sum_{\alpha > 0} z^{-\alpha -1} \Phi^+_{\alpha} \eqno\eq$$
$$\Phi_- (z) = \sum_{\beta > 0} z^{\beta -1} \Phi^-_{\beta} \eqno\eq$$

the resulting Hamiltonian  $H = H_+ + H_-$ gets the mode expansion:

$$H_+ = -\sum (2g \Phi^+_{\alpha + \alpha ' -2} \Pi^+_{\alpha '} +
\Phi^+_{\alpha  -2} \Phi^+_{\alpha - \alpha '}+ \Phi^+_{\alpha + \alpha ' -2}
\Phi^-_{\alpha '}) \Pi^+_{\alpha} \eqno\eq$$

where $[\Pi^{\pm}_{\alpha} , \Phi^{\pm}_{\alpha}] = \alpha \delta_{\alpha ,
\alpha' }$.

Let us end with several comments.
It is simple to extend the above derivations to higher critical models. For a
potential
 of power $Y^{k+1}$ one truncates the system to $k$ loop space fields
 $ \{ \Phi_0, \Phi_1 , \cdots \}$.Their splitting and rearrangement processes
 are given by $\Omega$ and $\omega$. Concerning these series of Hamiltonians
it may still be relevant to study the transition from the discrete to the
critical
continuum theory. For example, already
 in the above construction the expected $W_3$
structure or more specifically the $W_3$ generators are not manifestly visible.
Since they are certainly present in the S-D equations this deserves further
study.
\vskip .10in

\centerline{{\bf Conclusions}}

In the present work we have considered matrix models in the loop space
representation
and used them to construct field theories of  non-critical strings,
interpreting
the origin of time in terms of
stochastic quantization. Hamiltonians of Fokker-Planck type naturally arise.
They are represented
in loop space through collective field formalism.
The Hamiltonians in general take the form: $ H = - \sum_{i} O_i {d \over d
\rho_i}$
exhibiting a coupling between generators of an algebra $\{O_i \}$ and the
conjugate fields.
 Hermiticity of such Hamiltonians in general requires closure of the operator
algebra
and it follows that the partition function obeys the corresponding constraints.
The approach is exhibited giving simple string field theories for $d < 1$.
 The procedure of a continuum
limit is adressed. It is certainly not uninteresting to study this
correspondence
 between matrix models and string field theory. It might teach us about similar
correspondences in higher dimensions.

{\bf Note} During the writing of this paper we have received the work
referenced as [13].
There appears to be a considerable agreement between the two constructions.

{\bf Acknowledgements}

We are grateful to the Laboratoire de Physique Th\'eorique de l'ENS for their
hospitality while this work was done. We would like to thank Jean Avan for
discussions
and help,

\centerline{{\bf References}}

[1] E. Br\'ezin, V. Kazakov; Phys. Lett {\bf B236} (1990), 914.
D. Gross, A. Migdal; Phys. Rev. Lett. {\bf 64} (1990), 127.
M. Douglas, S. Shenker; Nucl.Phys. {\bf B335} (1991), 589.

[2] M. Fukuma, H. Kawai, R. Nakayama; Intern. Journ. Mod. Phys. {\bf A6}
(1991), 1385.
R. Dijkgraaf, E.Verlinde, H. Verlinde; Nucl. Phys {\bf B348} (19910, 435.

[3] S. Das, A. Jevicki; Mod. Phys. Lett {\bf A5} (1990), 1639.
J.Avan, A. Jevicki; Phys. Lett {\bf B266} (1991), 35.

[4] G. Moore, N. Seiberg, M. Staudacher; Nucl. Phys. {\bf B362} (1991), 665.

[5] I. Kostov; Nucl. Phys. {\bf B376} (1992), 539.

[6] N. Ishibashi, H. Kawai; KEK Preprint KEK-Th 364 (July 1993).

[7] H. Kawai, N. Kawamoto. T. Mogami, Y. Watabiki; Phys. Lett {\bf B306}
(1993), 19.

[8] G. Marchesini; Nucl. Phys. {\bf B239} (1984), 135.

[9] J. Rodrigues; Nucl. Phys, {\bf B260} (1986), 350.

[10] B. Sakita; Phys. Rev. {\bf D21} (1980), 1067.
A. Jevicki, B. Sakita; Nucl. Phys. {\bf B185} (1981),89.

[11] T. Tada, T. Yamaguchi; Phys. Lett {\bf B250} (1990), 38.
M. Douglas; Proceedings of Carg\`ese workshop 1990; NATO advanced Studies
Series {\bf B262}.
J.M. Daul, V.A. Kazakov, I.K. Kostov; Nucl. Phys. {\bf B409} (1993), 311.

[12] E. Gava, K. Narain; Phys. Lett. {\bf B 263} (1991), 213.
M. Staudacher; Phys. Lett {\bf B 305} (1993), 332.
J. Alfaro, J.C. Retamol; Phys. Lett. {\bf B222} (1989), 429.

[13] N. Ishibashi, H. Kawai; KEK preprint KEK-Th 378 ``String field theory of
$c \leq 1$ non-critical strings".
(december 1993).

\end

\end